\documentclass[a4paper,11pt]{article}
\pdfoutput=1 
\usepackage{jinstpub}

\usepackage[T1]{fontenc} 

\usepackage{graphicx}
\usepackage{amssymb,amsmath,array}
\RequirePackage{xspace}
\usepackage{hyperref}
\def\jpsi {\ensuremath{{J\mskip -3mu/\mskip -2mu\psi\mskip 2mu}}\xspace}
\mathchardef\Upsilon="7107
\def\Y#1S{\ensuremath{\Upsilon{(#1S)}}\xspace}
\def\etabOneS {\ensuremath{\eta_b{(1S)}}\xspace}

\def\Dbar    {\kern 0.2em\overline{\kern -0.2em D}{}\xspace}

\def\DD      {\ensuremath{D\Dbar}\xspace}
\def\Bbar    {\kern 0.18em\overline{\kern -0.18em B}{}\xspace}

\def\BB      {\ensuremath{B\Bbar}\xspace}

\def\CP {\ensuremath{C\!P}\xspace} 

\def\qqbar {\ensuremath{q\overline q}\xspace}
\def\uubar {\ensuremath{u\overline u}\xspace}
\def\ddbar {\ensuremath{d\overline d}\xspace}
\def\ssbar {\ensuremath{s\overline s}\xspace}
\def\ccbar {\ensuremath{c\overline c}\xspace}
\def\bbbar {\ensuremath{b\overline b}\xspace}

\def\epem {\ensuremath{e^+e^-}\xspace}

\newcommand{\kev}{\ensuremath{\mathrm{\,ke\kern -0.1em V}}\xspace}
\newcommand{\mev}{\ensuremath{\mathrm{\,Me\kern -0.1em V}}\xspace}
\newcommand{\mevcc}{\ensuremath{{\mathrm{\,Me\kern -0.1em V\!/}c^2}}\xspace}
\newcommand{\gev}{\ensuremath{\mathrm{\,Ge\kern -0.1em V}}\xspace}
\newcommand{\gevcc}{\ensuremath{{\mathrm{\,Ge\kern -0.1em V\!/}c^2}}\xspace}
\newcommand{\tev}{\ensuremath{\mathrm{\,Te\kern -0.1em V}}\xspace}
\newcommand{\tevcc}{\ensuremath{{\mathrm{\,Te\kern -0.1em V\!/}c^2}}\xspace}
\newcommand{\ev}{\ensuremath{\mathrm{\,e\kern -0.1em V}}\xspace}

\def\mum  {\ensuremath{{\,\mu\rm m}}\xspace}

\def\mmmrad{\ensuremath{\rm \,mm \cdot mrad}\xspace}  
\def\cms  {\ensuremath{{\rm \,cm}^{-2} {\rm s}^{-1}}\xspace}
\newcommand{\lum} {\ensuremath{\mathcal{L}}\xspace}

\newcommand{\EPEM}{\ensuremath{e^+e^-}\xspace}
\newcommand{\EMEM}{\ensuremath{e^-e^-}\xspace}
\newcommand{\GG}{\ensuremath{\gamma\gamma}\xspace}
\newcommand{\GE}{\ensuremath{\gamma e}\xspace}
\newcommand{\LGG}{\ensuremath{\mathcal{L}_{\gamma\gamma}}}

\newcommand{\LEE}{\ensuremath{\mathcal{L}_{ee}}\xspace}
\newcommand{\LEPEM}{\ensuremath{\mathcal{L}_{e^+e^-}}\xspace}
\newcommand{\WGG}{\ensuremath{W_{\gamma\gamma}}\xspace}

\newcommand{\CM}{\mbox{cm}\xspace}

\newcommand{\MKM}{\ensuremath{\mu m}\xspace}

\newcommand{\EN}{\ensuremath{\varepsilon_n}\xspace}

\newcommand{\be}{\begin{equation}}
\newcommand{\ee}{\end{equation}}
\newcommand{\bc}{\begin{center}}
\newcommand{\ec}{\end{center}}
\newcommand{\bi}{\begin{itemize}}
\newcommand{\ei}{\end{itemize}}
\newcommand{\ben}{\begin{enumerate}}
\newcommand{\een}{\end{enumerate}}

\title{\boldmath Gamma-gamma collider with $W_{\GG}\leq 12 \gev$ based on the 17.5 GeV SC linac of the European XFEL
}

\author[a,b]{V.~I.~Telnov}

\affiliation[a]{Budker Institute of Nuclear Physics,\\Novosibirsk, Russia}
\affiliation[b]{Novosibirsk State University,\\Novosibirsk, Russia}

\emailAdd{telnov@inp.nsk.su}

\abstract{We propose and demonstrate that a \GG\ collider with $\WGG \leq 12$ \gev can be added to the European XFEL with a minimal disruption to its main program.
High-energy photons will be obtained by Compton scattering of $0.5 \mum$ laser photons on the existing $17.5$ \gev electron beams.
Such a \GG\ collider would be an excellent place for the development and application of modern technologies: powerful lasers, optical cavities, superconducting linacs, and low-emittance electron sources --- as well
as training the next generation of accelerator physicists and engineers.
The physics program would include spectroscopy of $C=+$ resonances in various $J^P$ states ($b\bar{b}, c\bar{c}$, four-quark states, quark molecules and other exotica) in a mass range
barely scratched by past and not covered by any current or planned experiments.
Variable circular and linear polarizations will help in the determination of quantum numbers and measurement of polarization components of the \GG\ cross section ($\sigma_\perp$, $\sigma_\parallel$, $\sigma_0$, $\sigma_2$).}


\arxivnumber{2007.14003}

\begin{document}
\maketitle
\flushbottom

\section{Introduction}
Gamma-gamma collisions have a long history: since the 1970s, they have been studied at \epem storage rings in collisions of virtual photons ($\gamma^*$). Two-photon physics at storage rings is interesting and complementary to the main \EPEM physics program; however, it is not competitive because the number of virtual photons per electron is rather small: $\mathrm{d}n_\gamma \sim 0.03\; \mathrm{d}\omega/\omega$, therefore \LGG $\ll$ \LEPEM.
\begin{figure}[!htb]
\includegraphics[width=7.4cm]{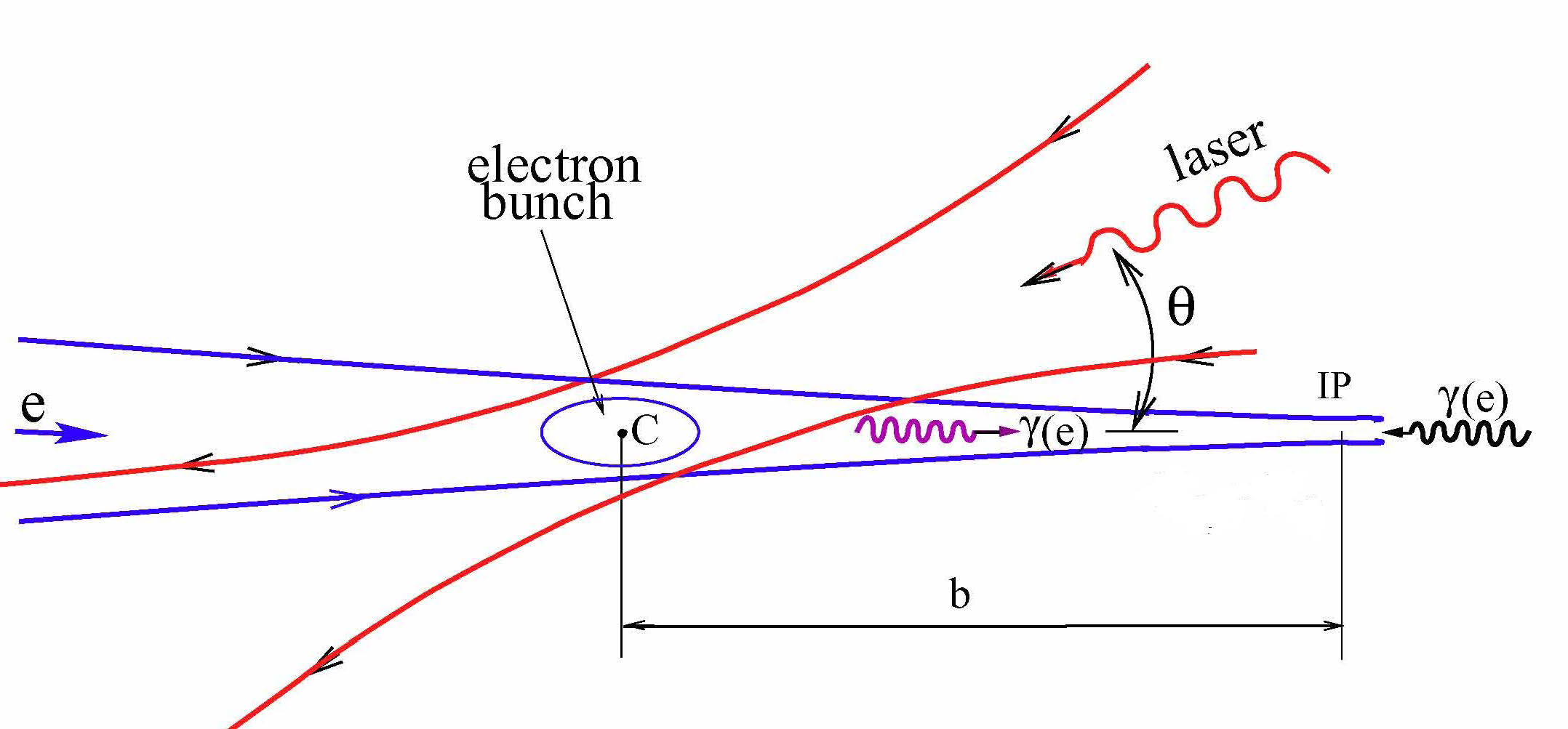} \includegraphics[width=7.4cm]{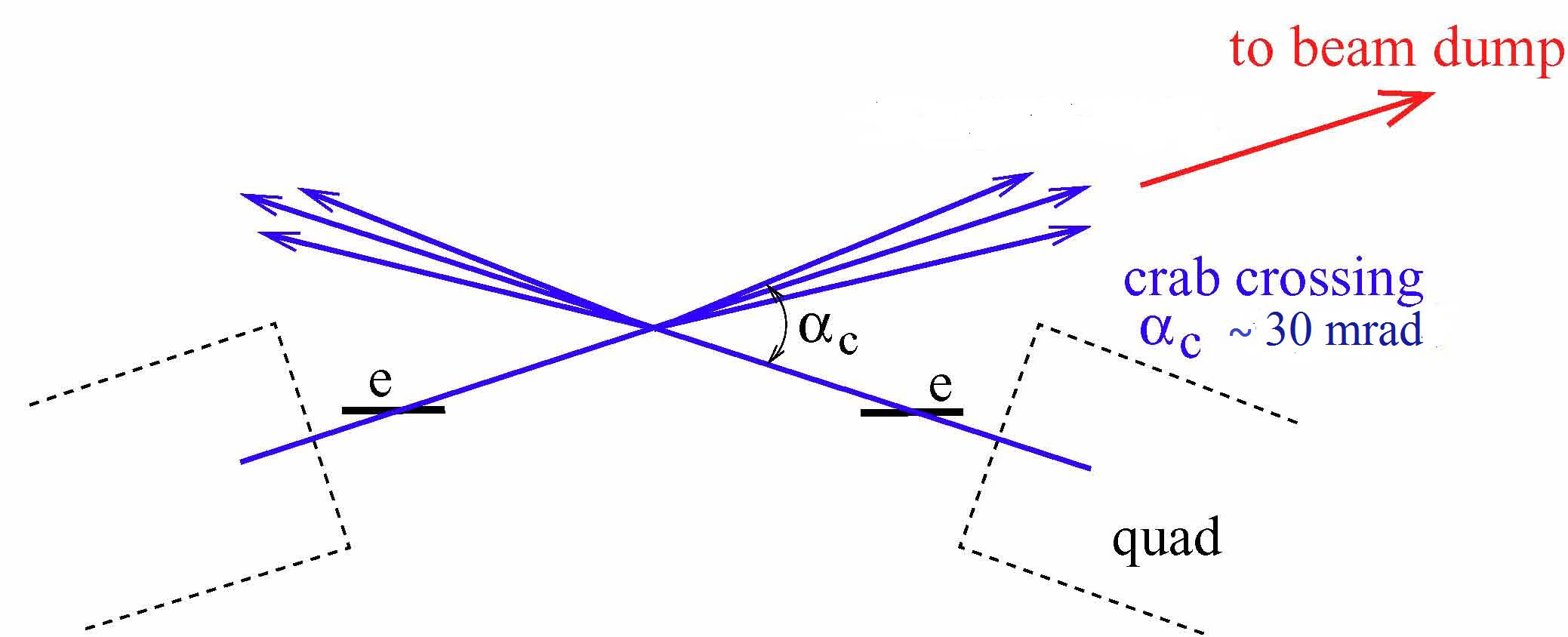}
\caption{\emph{(left)} The general scheme of a \GG, \GE photon collider. \emph{(right)} A crab-crossing collision scheme for the removal of disrupted beams from the detector to the beam dump. }
\label{scheme}
\end{figure}

At future \epem linear colliders beams will be used only once, which makes possible $e \to \gamma$ conversion by Compton backscattering of laser light just before the interaction point,
thus obtaining a \GG,\GE\ collider (a \emph{photon collider}) with a luminosity comparable to that in \EPEM\ collisions~\cite{GKST81,GKST83,GKST84}.
The general scheme of the photon collider is shown in Fig.~\ref{scheme}.
Laser photons scatter on electrons in the conversion region (CP) at the distance $b \sim \gamma \sigma_y$ from the interaction point (IP). If $\rho=b/\gamma \sigma_y \ll 1$, then at the IP photons of various energies are mixed in space and the invariant mass distribution of \GG\ system is broad.  In the opposite case, $\rho \gg 1$, the photons with higher energy collide at a smaller spot size and, therefore, contribute more to the luminosity and its spectrum becomes more monochromatic. After crossing the conversion region, electrons have a broad energy spectrum and large disruption angles due to deflection of low-energy electrons in the field of the opposing beam. The ``crab crossing'' collision scheme solves the problem of beam removal without hitting the final quadrupole magnets.

The maximum energy of scattered photons
\begin{equation}
\omega_m \approx \frac{x}{x+1}E_0; \;\;\;\;
x = \frac{4E_0\omega_0}{m^2c^4}
 \simeq 15.3\left[\frac{E_0}{\tev}\right]\left[\frac{\omega_0}{\ev}\right]=
 19\left[\frac{E_0}{\tev}\right]
\left[\frac{\mum}{\lambda}\right].
\label{x}
\end{equation}
For example: $E_0 = 250$ GeV, $\omega_0 = 1.17$ eV ($\lambda=1.06$ \MKM)  $\Rightarrow$ $x=4.5$
and $\omega_m/E_0 = 0.82$. So, the most powerful solid-state lasers with a 1 \mum wavelength are perfectly suited for a photon collider based on a $2E_0=$100--1000 \gev \epem linear collider, which have been actively developed since the 1980s (VLEPP, NLC, JLC, TESLA-ILC, CLIC). A \GG\ collider would require a laser with pulses of picosecond duration and an energy of a few joules, and a pulse structure similar to that of the \EMEM\ collider it is based on. The required laser system is within reach of modern laser technology.

Since the late 1980s, \GG\ colliders have been considered a natural part of all linear collider proposals; a number of detailed conceptual~\cite{NLC,TESLAcdr,JLC} and pre-technical designs~\cite{TESLATDR,telnov-acta2} have been published. With the existence of the Higgs boson confirmed, the photon collider is considered as one of the Higgs factory options~\cite{Telnov-higgs,Telnov-higgs-pc}; a dozen variants of stand-alone \GG\ Higgs factories have been proposed in addition to those based on ILC and CLIC. The advantages of a photon-collider Higgs factory include the somewhat lower beam energy required to produce the Higgs boson (compared to \epem collisions), no need for positrons and possibility to measure $\Gamma_{\GG}(H)$ more precisely than in \EPEM collisions. However, \epem colliders are better for the Higgs study overall due to the unique reaction $\epem \to ZH$ that allows detection of almost all Higgs decays, even invisible (by missing mass).

If a linear collider, either ILC or CLIC, is ever built, its first stage will offer only \epem collisions;  a photon collider is at least 30--40 years away. Obviously, such an outlook cannot inspire people who want to do something exciting and groundbreaking right now.

In April 2017, organizers of the ICFA Mini-Workshop on Future \GG\ Collider in Beijing, China assembled world experts in particle, laser and accelerator physics to discuss the current state of technology and brainstorm what can reasonably be achieved in this direction on a more relevant timescale. In my review talk, I proposed that a photon collider be built on the basis of the electron linac of an existing or future free-electron laser~\cite{pekin}. The first candidate is the European XFEL, which has been in operation since 2017. By pairing its 17.5 \gev electron beam with a 0.5 \mum laser, one can complement the European XFEL with a photon collider with a center-of-mass energy $W_{\GG} \leq 12$ \gev. While the $W_{\GG} <$ 4--5 \gev region can be studied at the \epem Super $B$-factory at KEK (in $\gamma^*\gamma^*$ collisions), in the $W_{\GG}=$ 5--12 \gev region this photon collider would have no competition. The addition of circular and linear polarizations would make such a photon collider a unique machine for the study of \GG\ physics in the \bbbar energy region, with many new states, including exotic, accessible for discovery and detailed study. This suggestion was further reported at several conferences~\cite{photon17,photon19}. The present paper is the first one on this subject.

\section{\boldmath Possible parameters of the \GG\ collider based on the European XFEL}
   The European XFEL has the following electron beam parameters~\cite{xfel}: beam energy $E_0=17.5$ \gev, number of particles per bunch $N = 0.62 \times 10^{10}$ (1 nC), bunch length $\sigma_z = 25$ \mum, normalized transverse emittance $\EN = 1.4 \mmmrad$, effective repetition rate 27 kHz (bunches are grouped into 10 trains per second, 2700 bunches per train, with an inter-bunch separation of 220 ns, or 66 m). To make a photon collider, the electron bunches from the XFEL should be alternately deflected into two arches with a radius of about 100 m and then converted to high-energy photons via Compton backscattering of laser light just before the interaction point (IP).

In order to transfer as much of the electron's energy as possible to the laser photon, the parameter $x$ should optimally be somewhat below $x=4.8$ ($\omega_m = 0.82\, E_0$), or $\lambda = 4.2\, E_0[\tev]\,\mum$~\cite{TESLATDR}. For $E_0=17.5 \gev$, $\lambda \approx 0.074 \mum$ and the maximum $W_{\GG} \approx 0.8 \times 2 E_0 = 28 \gev$. We do not consider this extreme case for two reasons: 1) a laser system with such a short wavelength and the required peak (terawatt) and average (100 kW) power is technically infeasible; 2) there is no interesting physics in the $W_{\GG}$ region between $12 \gev$ and $28 \gev$.
Instead, we consider the laser wavelength $\lambda=0.5\mum$, having in mind a laser system with an external optical cavity, as proposed for the ILC-based photon collider~\cite{telnov-acta2}, pumped by a frequency-doubled $1 \mum$ laser. In this case, $x=0.65$ and $\omega_m/E_0=x/(x+1) \approx 0.394$. The laser intensity in the conversion region is limited by nonlinear effects in Compton scattering, described by the parameter $\xi^2$ (see ref.\cite{TESLATDR}). We assume $\xi^2=0.05$, which reduces $\omega_m/E_0$ by $3\%$, down to 0.38. In this case, $W_{\GG, \rm max}=13.3 \gev$, with a peak at $12 \gev$, which covers the region of \bbbar resonances. The peak energy can be varied by the electron beam energy.  In the case of $\lambda = 1 \mum$, the peak  energy is 7.5 GeV for $2E_0=35 \gev$, this option is technically easier and can be used for $W_{\GG} \leq 7 \gev$.

In calculating the optimal laser pulse duration and flash energy, we assume a laser system similar to that proposed for the ILC-based \GG\ collider (optical resonator, mirrors outside electron beams)~\cite{telnov-acta2}. The thickness of the laser target is taken to be equal to one scattering length for electrons with an initial energy of 17.5 GeV. The required flash energy is about 3 J, that is smaller by a factor of 3 than in the ILC case thanks to the larger Compton cross section at smaller values of $x$.

One of the most serious issues in \GG\ collider design is the removal of used beams, which are disrupted by the opposing electron beam. The disruption angle for low energy electrons is proportional to $\sqrt{N/\sigma_z}$. In order to keep disruption angles within an acceptable range, we assume the bunch length longer than at the XFEL, 70 \mum instead of 25 \mum.

Simulation of the processes at the conversion and interaction points was performed with my code that has been in development since 1995 and used to optimize photon collider designs for the NLC, CLIC and TESLA-ILC~\cite{TESLATDR}. We consider both unpolarized (as currently available at the European XFEL) and $80\%$ longitudinally polarized electron beams. The laser beam should be circularly polarized, $P_c= \pm 1$, when circularly polarized high-energy photons are needed. Collisions of linearly polarized photons would also be of interest for physics; for that, linearly polarized laser beams should be used. The degree of circular polarization in the high-energy part of spectrum can be close to $100\%$ (for any $x$)  and about $85\%$ for linear polarization (for $x=0.65$)~\cite{TESLATDR}.

\begin{figure}[!htb]
\centering
\includegraphics[width=5.5cm,height=5.5cm,angle=0]{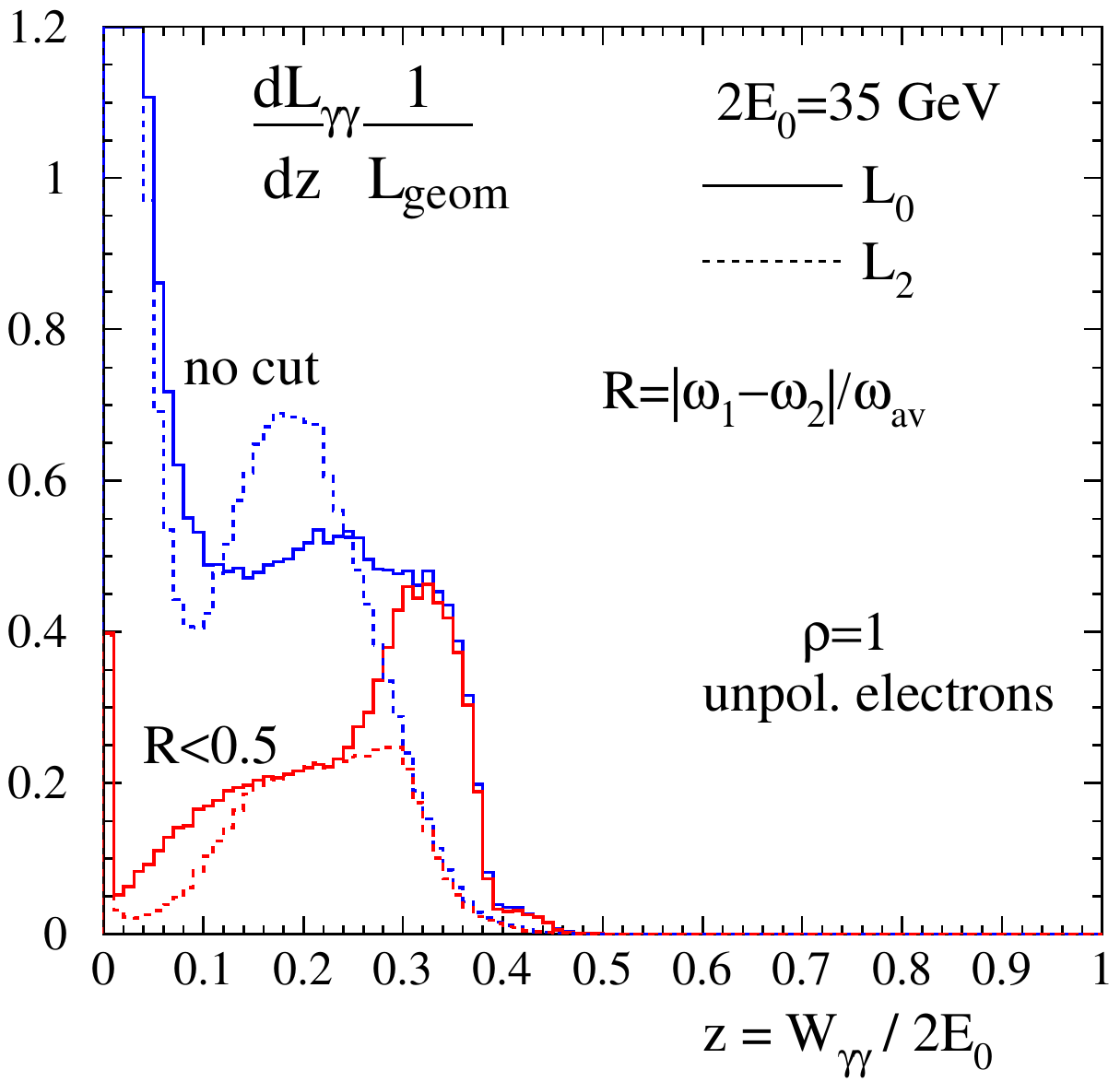} \hspace*{1cm} \includegraphics[width=5.5cm,height=5.5cm,angle=0]{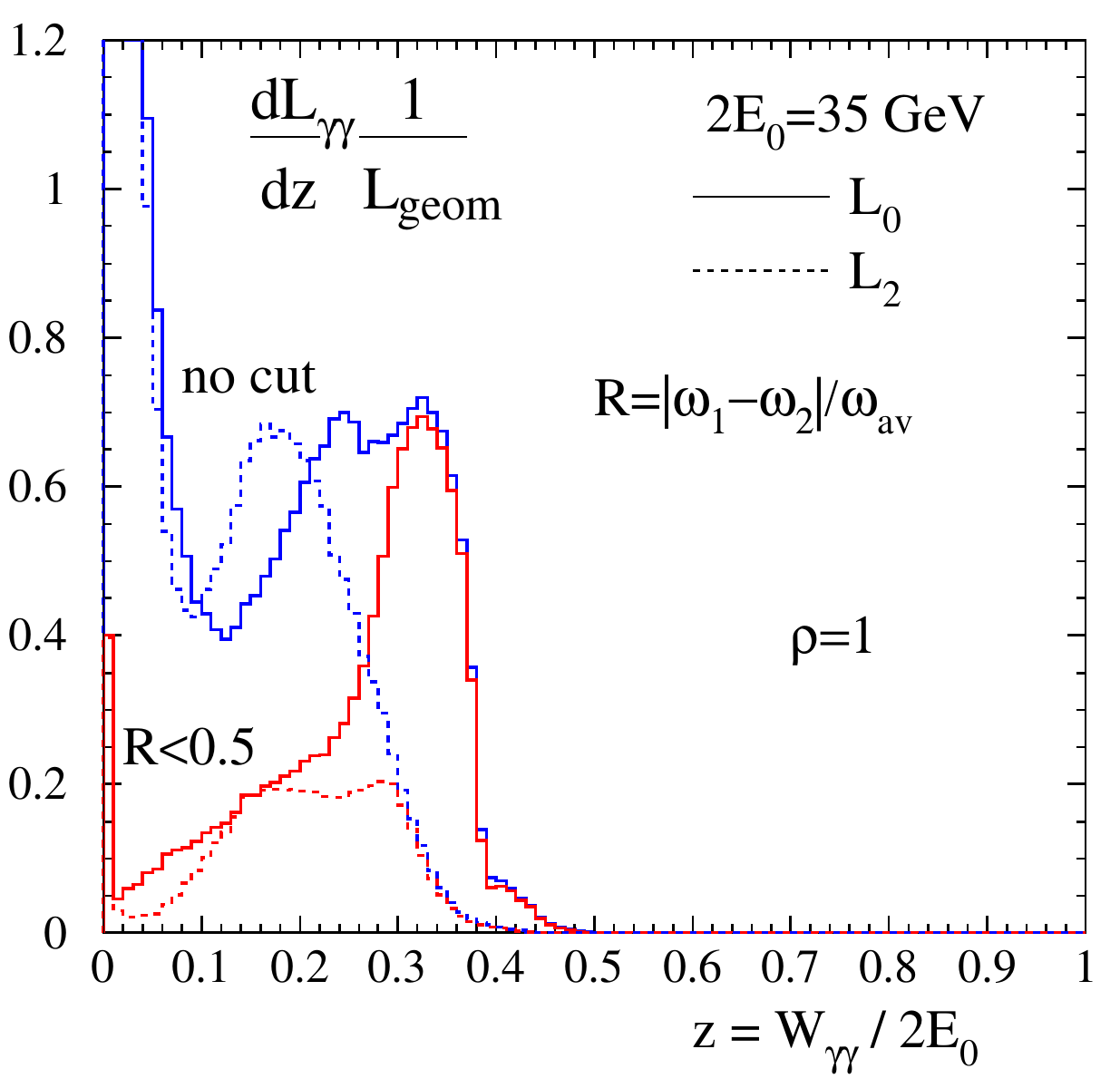}
\vspace*{-0.2cm}
\caption{\GG\ luminosity distributions vs the invariant mass $W_{\GG}$: \emph{(left)} unpolarized electrons; \emph{(right)} longitudinal electron polarization $2\lambda_e=0.8$ (80$\%$). In both cases the laser photons are circularly polarized, $P_c=-1$. Solid lines are for the total helicity of the two colliding photons $J_z=0$, dotted lines for $J_z=2$. Red curves are luminosities with a cut on the longitudinal momentum.} \label{pol-np}
\end{figure}

\begin{figure}[!htb]
     \begin{center}
\includegraphics[height=4.8cm,clip=true] {gg-35-pol-1.pdf} \includegraphics[height=4.8cm,clip=true] {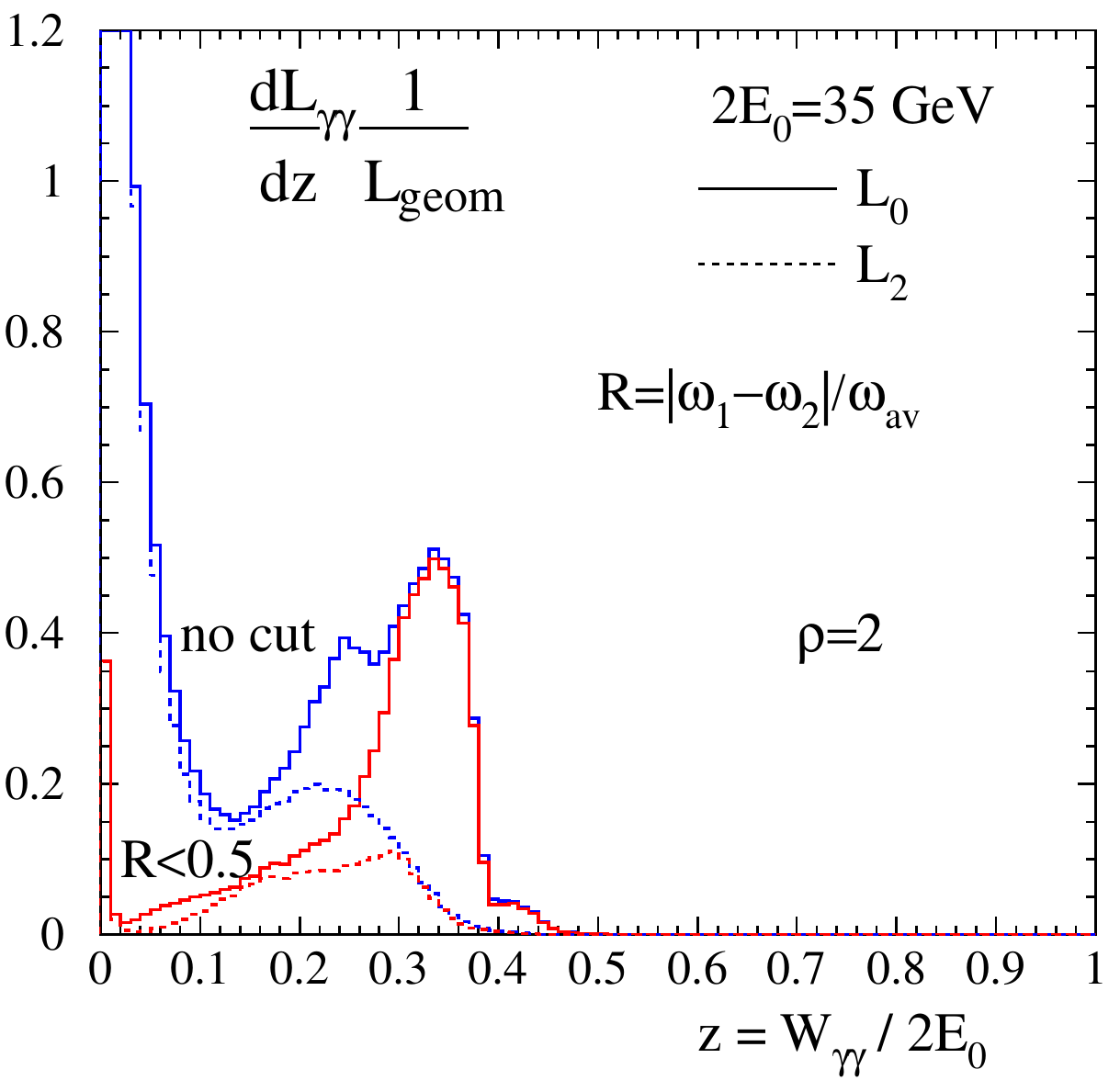}
\includegraphics[height=4.8cm,clip=true] {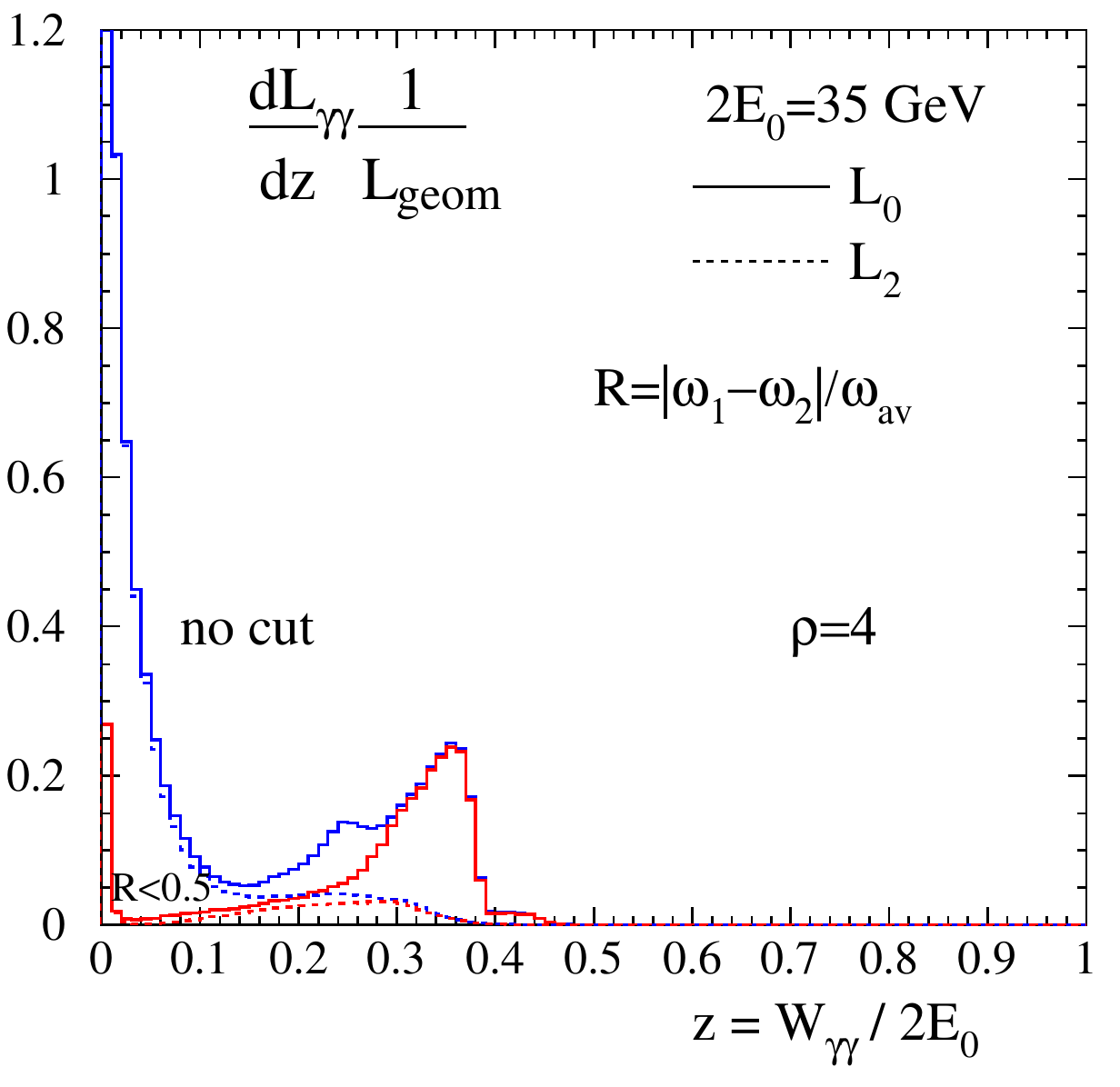}
       \vspace*{-0.7cm}
     \end{center}
     \caption{\GG\ luminosity distributions for various distances between the conversion and interaction points characterized by the parameter $\rho=b/(\gamma \sigma_y)$. See also comments in the text and Fig.~\ref{pol-np}.}

   \vspace*{-0.cm}
   \label{lum-rho-1}
   \end{figure}

\begin{figure}[!htb]
     \begin{center}
\includegraphics[height=4.9cm,clip=true] {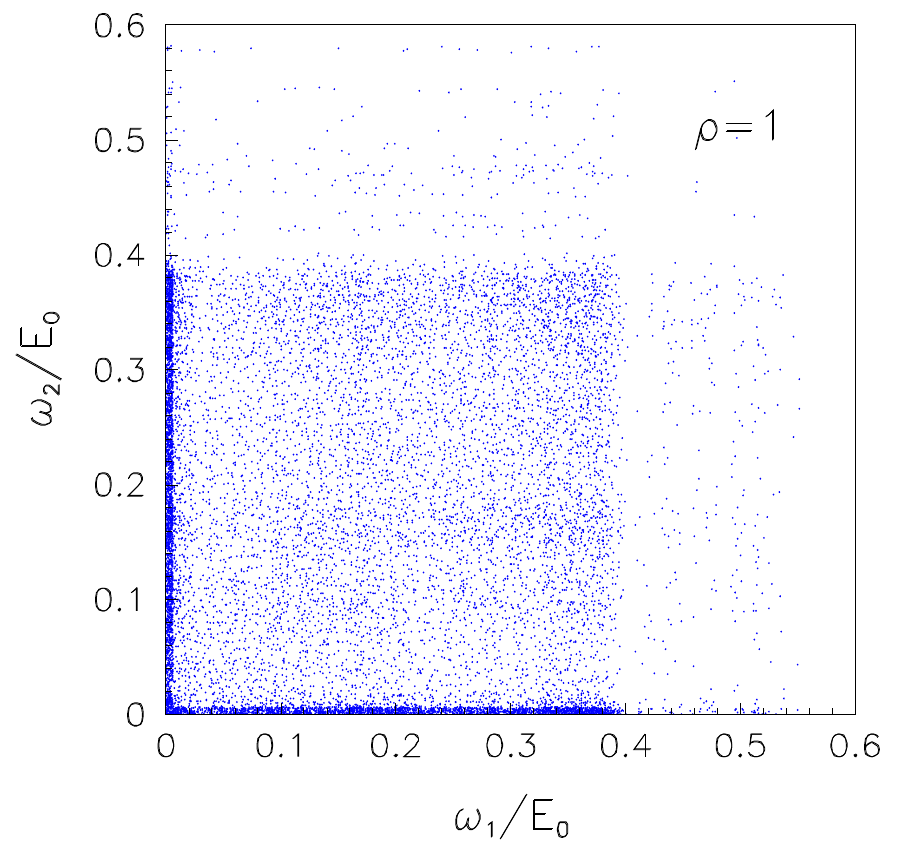} \hspace{-0.5cm} \includegraphics[height=4.9cm,clip=true] {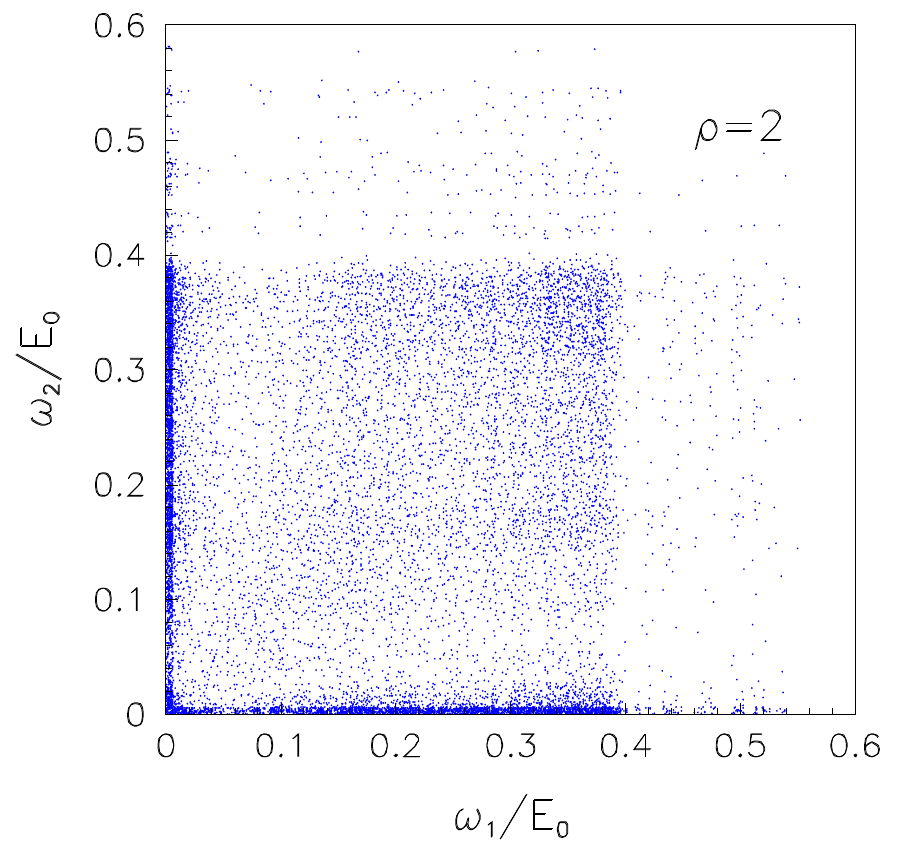} \hspace{-0.5cm}
\includegraphics[height=4.9cm,clip=true] {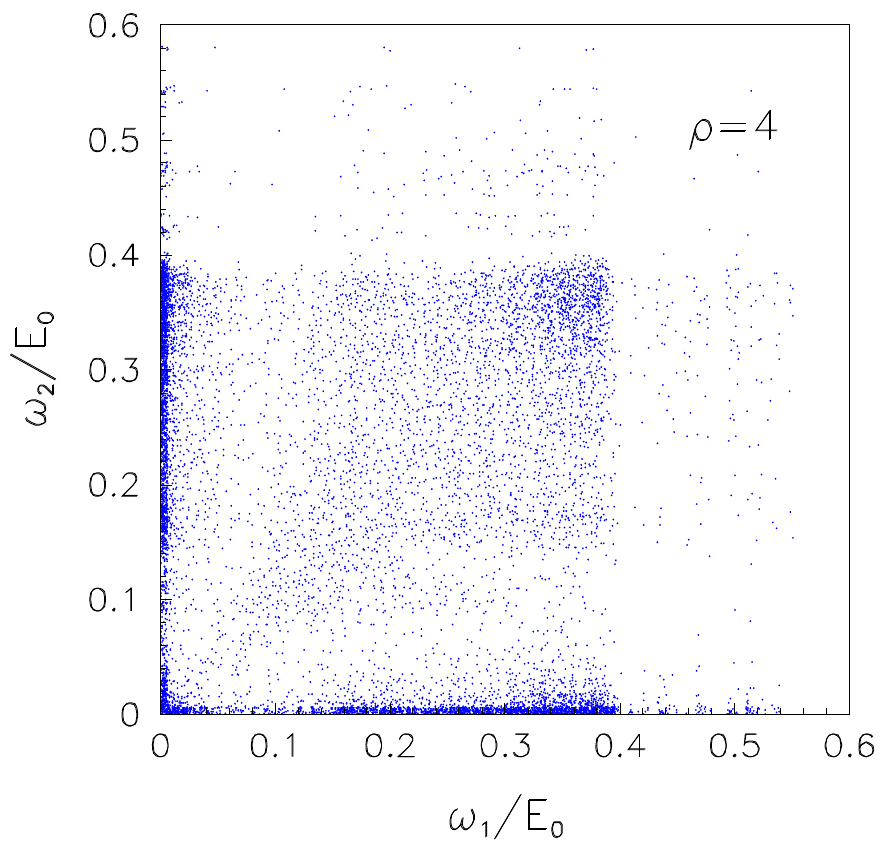}
       \vspace*{-0.6cm}
     \end{center}
    \caption{\GG\ luminosity plots vs energies of colliding photons $\omega_1$ and $\omega_2$  for various distances  between the conversion and interaction points. See comments in the text and Fig.~\ref{lum-rho-1}. }
   \vspace*{-0.2cm}
   \label{lum-rho-2}
   \end{figure}

The \GG\ luminosity spectra for non-polarized and longitudinally polarized electrons are shown in Fig.~\ref{pol-np}.
The spectra are decomposed into states with the total helicity  of the colliding photons $J_z=0$ or $2$; the total luminosity is the sum of the two spectra.
Also shown are the luminosities with a cut on the relative longitudinal momentum of the produced system that suppresses boosted collisions of photons with very different energies. Luminosity distributions similar to those in Fig.~\ref{pol-np} but for various distances between the conversion and interaction points are shown in Fig.~\ref{lum-rho-1}: as $\rho$ increases, the luminosity spectra become more monochromatic at the cost of some reduction in luminosity. Fig.~\ref{lum-rho-2} shows the \GG\  luminosity distribution vs energies of coliding photons $\omega_1$ and $\omega_2$ for the same conditions as in Fig.~\ref{lum-rho-1}. Beside Compton photons one can clearly see here low energy photons from beamstrahlung. Rare photons with $\omega/E_0>0.38$ are due to nonlinear effects in Compton scattering, when one electron scatters simultaneously on several laser photons. Main interest present collisions of high energy photons with good polarization properties. For measuring total \GG\ cross sections one should work with $\rho \sim$ 3--4, when luminosity spectrum is more monochromatic. For study of resonances, when the invariant mass is determined by the detector, the maximum luminosity is needed, therefore $\rho < 1$ is preferable. The resulting photon collider parameters are given in Table~\ref{Table1}.

 The \GG\ luminosity is proportional to the geometric $ee$-luminosity which is determined by the transverse beam emittances. At the European XFEL electrons are generated by RF gun with solid photocathode and emittances are limited by space charge effects. On the other hand, there has been significant progress in the development of electron sources with a plasma cathode. Here the accelerating fields are several orders of magnitude higher, which allows one to bypass the space charge limit. Plasma accelerator experts assure that  it is possible to make a source of (unpolarized) electrons for the photon collider discussed here with the emittance about 15 times smaller than that in the European XFEL injector, which will lead to an increase of the luminosity by the same factor!


 \begin{table}[!hbtp]
 \caption{Parameters of the proposed photon collider based on the European XFEL.}
{
\renewcommand{\arraystretch}{0.85} \setlength{\tabcolsep}{2mm}
\begin{center}
\begin{tabular}{| l | l | c |  }  \hline
$2E_0$&GeV&35 \\
$N$ per bunch & $10^{10}$ & 0.62 \\
Collision rate & kHz & 13.5 \\
$\sigma_z$ &$\mum$& 70 \\
$\varepsilon_{x,\,n}/\varepsilon_{y,\,n}$ & \mmmrad & 1.4/1.4 \\
$\beta_x/\beta_y$ at IP & $\mum$ & 70/70 \\
$\sigma_x/\sigma_y$ at IP & nm & 53/53 \\
Laser wavelength $\lambda$ & $\mum$ & 0.5 \\
Parameters $x$ and $\xi^2$ & & 0.65, 0.05 \\
Laser flash energy & J & 3 \\
Laser pulse duration & ps & 2 \\
f$\# \equiv F/D$ of laser system & & 27 \\
Crossing angle & mrad & $\sim 30$ \\
$b$ (CP--IP distance) & mm & 1.8 \\
${\LEE}_{\rm ,geom}$ & $10^{33}$\cms & 1.45 \\
$\LGG\,(z>0.5z_m)$ & $10^{33}$\cms & 0.19 \\
$W_{\GG}$ (peak) & GeV & 12 \\
\hline
\end{tabular}
\end{center}
\vspace{-0mm}
}

\label{Table1}
\end{table}

\section{\boldmath Physics at a $0.1$--$12$\gev photon collider}

The photon, like the electron, is a point-like particle that participates in electromagnetic interactions.
For $\WGG \gg mc^2$ and $|cos{\theta}|<0.9$, the cross section for fermion-pair production in \GG collisions (unpolarized) is $\sigma_{\GG} \approx 0.5(q/e)^4/\WGG^2[\gev^2] \cdot 10^{-30}\; \CM^2$, while in \EPEM\ collisions in the same range of angles $\sigma_{\EPEM} \approx 0.075 \,(q/e)^2/W_{\EPEM}^2[\gev^2] \cdot 10^{-30}\; \CM^2$.

Additionally, a photon spends some time in the form of virtual charged-fermion pairs, and thus can behave as a hadron (virtual vector meson) in \GG\ collisions.
The cross section $\sigma(\GG \to \mathrm{hadrons})=$(0.4--0.6)$\times 10^{-30}\; \CM^2$ for $W_{\GG}>1\gev$; note that it does not decrease with energy and is significantly greater than the point-like quark-pair production cross section.

A nice feature of both \EPEM and \GG\ collisions is the single resonance production of hadrons. At \EPEM\ colliders, resonances with the photon quantum numbers, $J^{PC}=1^{--}$, can be single-produced, which includes the \jpsi and $\Upsilon$ families. On the other hand, two real photons can single-produce $C=+$ resonances with the following quantum numbers~\cite{Landau}: $J^P=0^+$, $0^-$, $2^+$, $2^-$, $3^+$, $4^+$, $4^-$, $5^+$, etc., the forbidden numbers being $J^P=1^{\pm}$ and (odd $J)^-$. Therefore, the \GG\ collider presents a much richer opportunity for the study of hadronic resonances.

Resonance production cross sections in \GG\ collisions depend on the total helicity of the two photons, $J_z=0$ or 2. Assuming that the $C$ and $P$ parities are conserved, resonances are produced only in certain helicity states~\cite{Landau}: $J_z=0$ for $J^P=0^\pm$, (even $J)^-$; $J_z=2$ for (odd $J \neq 1)^+$; $J_z=0$ or 2 for $J^P$ = (even $J)^+$. In the experiment, the value of $J_z$ is chosen by varying the laser photon helicities.

Photon polarization is characterized by the photon helicity $\lambda_\gamma$, the linear polarization $l_\gamma$, and the direction of the linear polarization. Any \GG\ process is described by 16 partial cross sections, but only three are important as they do not vanish after averaging over the spin states and azimuthal angles of the final particles~\cite{GKST84,Pak};
they are $\sigma^{np}=(\sigma_\parallel+\sigma_\perp)/2 = (\sigma_0+\sigma_2)/2$, $\tau^c=(\sigma_0-\sigma_2)/2$, and $\tau^l=(\sigma_\parallel-\sigma_\perp)/2$.

The number of events
\be
\mathrm{d}\dot{N} = \mathrm{d}\LGG ( \mathrm{d}\sigma^{np} + \lambda_{\gamma} \tilde{\lambda}_{\gamma}\;
\mathrm{d}\tau^c +   l_{\gamma} \tilde{l}_{\gamma} \cos{2\Delta\phi} \;\mathrm{d}\tau^l)\;,
\ee
where the tilde sign marks the second colliding beam and $\Delta\phi$ is the angle between the directions of the linear polarizations of the two colliding photons.
For example, for a $J=0$ resonance, $\sigma_2$ is always zero, while $\sigma_\parallel$ and $\sigma_\perp$ depend on \CP parity: for a $\CP=1$ resonance, $\sigma_\parallel=\sigma_0$, $\sigma_\perp=0$; for $\CP=-1$, $\sigma_\parallel=0$, $\sigma_\perp=\sigma_0$. The cross section in this case ($J=0$)
\be
\sigma \propto 1 + \CP \cdot l_{\gamma,1}l_{\gamma,2}\cos{2\Delta\phi}.
\ee
Scalar particles are produced when linear polarizations of the colliding photons are parallel; pseudoscalars are produced when the polarizations are orthogonal.
Therefore, availability of circular and linear photon polarizations is of great value in the determination of $J^P$ of the produced resonances and allows one to measure all important polarization components of the \GG\ cross sections.

Photon colliders have broad luminosity spectra with complicated polarization properties. All these characteristics can be measured and calibrated experimentally using QED processes with known cross sections~\cite{TESLATDR,Pak}.

At a \GG\ collider with $ W < 12 \gev$ (i.e., operating in the \uubar, \ddbar, \ssbar, \ccbar, \bbbar energy region), a high degree of circular and linear photon polarization is available in the high energy part of the luminosity spectrum (see formulas and graphs in ref.~\cite{TESLATDR}). The degree of circular polarization $\lambda_{\gamma}$ is almost 100\% and is determined by the laser polarization. The degree of linear polarization at $\omega=\omega_m$ is $l_{\gamma}=2/(1+x+(1+x)^{-1})$, that is 0.88 for $x=0.65$.
 Longitudinal polarization of the electron beam is desirable as it enables enhancement of photon helicities and making larger the ratio $\lum_0/\lum_2$ (or its opposite), see Fig.~2 (right).

Observation of \GG\ resonances is one of the most interesting tasks for the proposed \GG\ collider. The production cross section for a resonance is proportional to its partial width $\Gamma_{\GG}$, which says a lot about its structure and nature. While nearly all of the observed \GG resonances are \qqbar states, there are also several candidates for four-quark states. Glueballs with $C=+$, composed of two gluons, have been predicted but not observed yet.

Particles with $C=+$ are observed at \EPEM\ colliders in decays of heavier particles, such as \jpsi, $\Upsilon$ and their excited states. Decay branchings are not small only in decays of narrow states. The \jpsi ($\Upsilon$) excited states with masses above the \DD (\BB) threshold are broad; therefore, their branching to $C=+$ states are very small. A photon collider would allow not only to produce $C=+$ states directly but also simultaneously to measure their $\Gamma_{\GG}$.

For example, the $C=+$ meson \etabOneS, the ground state of the \bbbar system, has been observed at $B$-meson factories in radiative decays of \Y2S and \Y3S;
however, its two-photon partial width $\Gamma_{\GG}$ remains unknown because the branching fraction $\etabOneS \to \GG$ is expected to be less then $10^{-4}$.
At the photon collider, the production rate of a resonance with $J=0$ is
   \be
      \dot{N} = \frac{\mathrm{d}\LGG}{\mathrm{d}W_{\GG}}\frac{4\pi^2\Gamma_{\GG}(1+\lambda_1\lambda_2)(\hslash c)^2}{(Mc^2)^2} \approx 8 \times 10^{-27}\frac{\Gamma_{\GG} \LEE}{E_0 M^2[\gev^2]},
   \ee
where we substitute $(\mathrm{d}\LGG/\mathrm{d}W_{\GG})(2E_0/\LEE) \approx 0.5$ (see Fig.\ref{pol-np} for unpolarized electrons) and photon helicities $\lambda_{1,2}=1$.
For  $E_0=17.5 \gev$, $\Gamma_{\GG}(\eta_b)=0.5 \kev$, $M_{\eta_b}=9.4\gev$, $\LEE=1.45\times 10^{33} \cms$ and $t=10^7$ s, we would get 37500 events. With plasma electron injector, there could be over 0.5 million $\eta_b$ events.  Electron polarization, if available, would increase the \etabOneS production rate by a factor of 1.5.

While the LEP2 collider did have enough energy to produce $\eta_b$ in $\gamma^*\gamma^*$  collisions, it was not observable because its production rate at \LEE(LEP2) $=10^{32} \cms$ was about 1000 times lower than at the \GG\  collider proposed herein (without plasma source). In order to have the same production rate of $\GG$ states in the central region ($0.2<\omega_1/\omega_2 <5$), an \EPEM\ collider with $2E_0 \sim$ 100--200 \gev would need a luminosity $\LEPEM \sim 10^{35}$, or a factor 70 greater than the geometric luminosity $\LEE \sim 1.45 \times 10^{33}$ at the proposed \GG\ collider based on the European XFEL.

Observation of single-produced $C=+$ resonances in \GG\ collisions requires the detection of all final-state particles, which can be checked by requiring that the total transverse momentum be near zero. These events will be central, with a more-or-less isotropic distribution of particles. The non-resonance hadronic background is large but can be suppressed using a cut on $\sum|p_{i,_\perp}|$ and particle identification ($b,c$-quark tagging).
Also note that at this photon collider only the high-energy parts of the luminosity spectra have good polarization properties. In order to use these good properties over the entire energy range, one has to do an energy scan by operating at different electron beam energies.

\section{Conclusion}

Photon colliders are highly cost-effective additions to future \EPEM\ linear colliders.
Unfortunately, the outlook for high-energy linear colliders has been uncertain for many decades. It therefore makes sense to build a photon collider for a smaller energy, $W_{\GG}<12 \gev$, which would offer excellent coverage of the \bbbar and \ccbar production regions. The \GG\ physics in this energy range is very rich, much of it not accessible otherwise. The required linear accelerator already exists: it is the superconducting $17.5 \gev$ linac of the European XFEL. The photon collider can use the "spent" electron beams, which currently are sent to the beam dump (although for some experiments time sharing would be desirable). This \GG\ collider would be a nice place for the application and further development of cutting-edge accelerator and laser technologies. It does not need positrons or damping rings. The required laser system is identical to that needed for the photon collider at the ILC. While it cannot be guaranteed that the proposed \GG\ collider would yield any breakthrough discoveries (which applies to all other projects as well), there are many arguments (scientific, technical, financial and social) in favor of such a collider of a brand new type.

\acknowledgments
  This work was supported by RFBR-DFG Grant No 20-52-12056. I would like to thank W.~Chou, Wei~Lu, C.~Zhang and other organizers of ICFA Mini-Workshop on Future gamma-gamma Collider in Beijing, which stimulated appearance of this proposal. I am very grateful to D.~Asner, C.~Barty, R.~Brinkmann, S.~Brodsky, R.~Byer,  A.~de-Roeck, I.~Ginzburg, J.~Gronberg, R.~Heuer,  K.~J.~Kim, G.~Korn, K.~M$\ddot{\rm o}$nig, G.~Moortgat-Pick, G.~Mourou, I.~Pogorelsky, F.~Richard, V.~Serbo, T.~Takahashi, M.~Velasco, V.~Yakimenko, K.~Yokoya, A.~F.~Zarnecki  for interest to future \GG\ colliders and useful discussions.

\end{document}